\documentclass{vgtc}                          

\ifpdf
  \pdfoutput=1\relax                   
  \usepackage{graphicx}                
  \DeclareGraphicsExtensions{.pdf,.png,.jpg,.jpeg} 
\else
  \usepackage{graphicx}                
  \DeclareGraphicsExtensions{.eps}     
\fi%

\graphicspath{{figures/}{pictures/}{images/}{./}} 

\usepackage{microtype}                 
\PassOptionsToPackage{warn}{textcomp}  
\usepackage{textcomp}                  
\usepackage{mathptmx}                  
\usepackage{times}                     
\usepackage{cite}                      
\usepackage{tabu}                      
\usepackage{booktabs}                  
\usepackage[skip=0pt]{caption}

\onlineid{0}

\vgtccategory{Research}

\vgtcinsertpkg

\title{Animated Drag and Drop Interaction for Dynamic Multidimensional Graphs}

\author{Benjamin Renoust\thanks{e-mail: renoust@ids.osaka-u.ac.jp}\\ %
    \parbox{1.8in}{\scriptsize \centering Osaka University Institute for Datability Science (IDS) \& National Institute of Informatics (NII)}
\and Haolin Ren\thanks{e-mail: haolin.ren@ina.fr}\\ %
     \parbox{1.8in}{\scriptsize \centering National Audiovisual Institute (INA) \& University of Bordeaux \\ LaBRI CNRS UMR 5800} %
\and Guy Melan\c{c}on \thanks{e-mail: melancon@labri.fr}\\ %
     \parbox{1.8in}{\scriptsize \centering University of Bordeaux \\ LaBRI CNRS UMR 5800}}

\abstract{In this paper, we propose a new drag and drop interaction technique for graphs. We designed this interaction to support analysis in complex multidimensional and temporal graphs. The drag and drop interaction is enhanced with an intuitive and controllable animation, in support of comparison tasks.%
} 

\begin{document}



\maketitle

\section{Introduction} 
When facing multidimensional datasets, analysts working on non-trivial tasks systematically needs to jump between facets of their data \cite{becker1987dynamic, elmqvist2008rolling}.  When facets could be different representations on the data, they may also encapsulate different aspects of the data, under conditional filters \cite{wilhelm2007linked}. This brings obviously the necessity to put side by side different views (\emph{i.e.} different representations of a same dataset), thus using in analysts the best of human's perception to spot and understand the differences or similarities in pattern \cite{koffka1922perception}. 

\begin{figure*}[h]
\centering
    \includegraphics[width=1\textwidth]{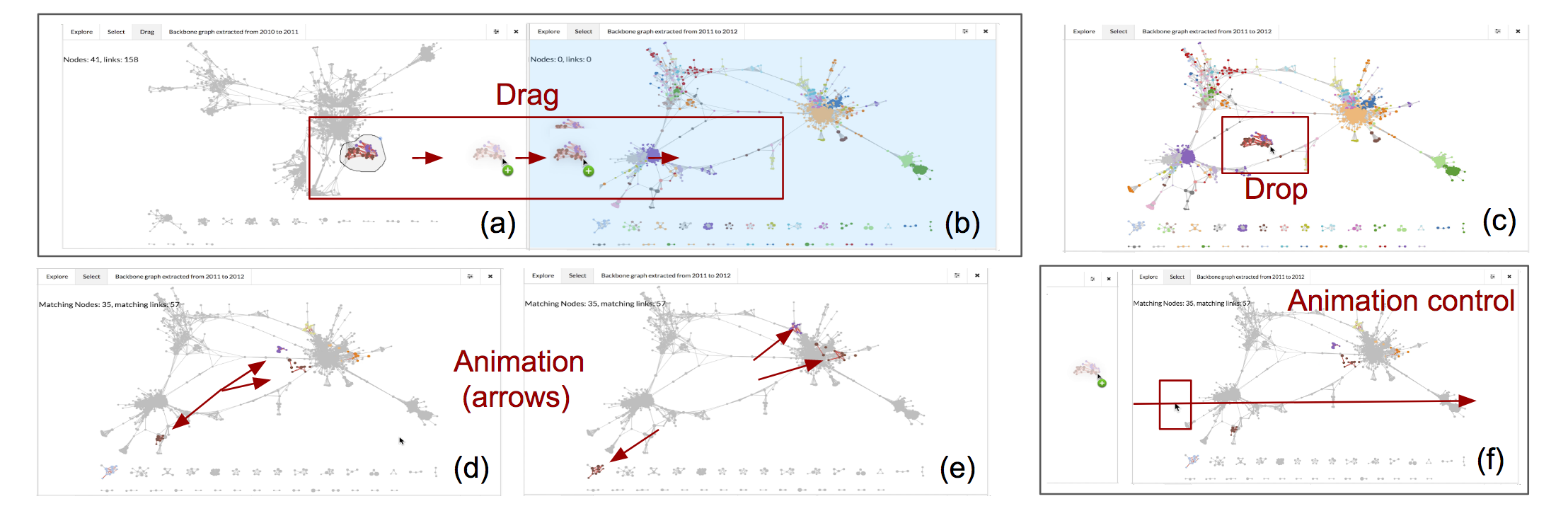}
    \caption{User select a subgraph from a view (a), then drag the selected subgraph to another view (b), then drop the subgraph (c). The same nodes and links will gradually move to the matching location and then be highlighted (d-e). User may go backward or forward in the animation holding a control key and positioning the mouse along an invisible line between the views (f). \vspace{-0.3cm}
}
    \label{fig:drag-drop}
\end{figure*}

It became completely natural to wish to find the same elements appearing in one view within another view. This requires interaction design, such that users may further search individual datum or groups of data. The multiple views with the interaction operating between them became known as \emph{linked views} \cite{wills2007linked}. They became a paradigm early on in visualization \cite{becker1987dynamic} as well as most standard user interfaces from OS \cite{apple1992macintosh} to softwares \cite{auber2017tulip}.

The archetype of linked views certainly is represented by the coordinated scatter plot matrix \cite{becker1987dynamic}, and this type of selection$+$highlight coordination (\emph{a.k.a} \emph{linked highlighting}) later became standard in \cite{elmqvist2008rolling, heinrich2012parallel}. It naturally combines with the decomposition of complex datasets in small multiple views (or small multiples \cite{tufte1990envisioning}) and linked-highlighting became essential to all sort of manipulations in complex graphs whether from dimensionality \cite{renoust2015detangler}, or temporality \cite{renoust2016visual}, or even combined \cite{ren2018generating}.

\section{Temporal Multidimensional Graphs}

We consider a general class of graphs that captures a high level of complexity that we will refer to as temporal multidimensional graphs ${\cal{G}}=(V,E,A,T)$ (weights and directions do not matter here). Our graphs are also multidimensional because links and nodes contain attributes. Our graphs are temporal such as nodes and links exist over time. This family of graphs spawns from real world complex datasets, for which comparison tasks are essential, so we put ourselves in a setting of small multiples, each view defined by time or by other properties. 

The small multiples approach has demonstrated to be a good strategy for topology-based tasks on dynamic graphs \cite{archambault2011animation}. In contrast, animation performs well for tasks related to the dynamic evolution of a graph such as addition and deletion of nodes and edges \cite{archambault2011animation}. The use of animation is controversial, but animating transitions, especially with the support of interaction, may be useful \cite{tversky2002animation}. 

\section{Coordinating Temporal Multi Graphs}

We place our work in a framework, where multiple subgraphs are defined, and each subgraph is coordinated with each other with linked highlighting. One of our tasks is to investigate the evolution of communities formed over time. These communities form around nodes and edges that exist in different time frames, but not necessary exist persistently across \emph{all} timeframes: nodes and links may appear, disappear, and nodes may even switch community.

Linked highlighting across small multiples brings forward a selection in one view to all other views simultaneously. When users select nodes and links in a view, these appear in all other views. Although very convenient to dispatch a selection to many other views and understand a global trend, it is hard to follow which elements goes where. A task better suited for animated transitions. 

\subsection{A Drag and Drop Interaction for Graphs} 

As an alternative, we propose a drag and drop interaction (Fig. \ref{fig:drag-drop}) which proposes a trade-off between the advantages of animation and those of small multiples with linked highlighting. The principle of this drag and drop interaction is to allow user to physically grab a part of their graph and observe how it arranges in a different view.

This interaction fits into the linked view paradigm \cite{wills2007linked}, as we have the three elements necessary to its execution. \emph{1) A view capable of interaction:} the view from which we grab a subgraph. \emph{2) A mechanism to transfer information:} the drag that physically transfers a subgraph. \emph{3) A view capable of visually respond:} the animation occurring upon release.

More specifically, each view among the small multiples displays one subgraph $G_i \subseteq {\cal{G}}$. Similarly to linked-highlighting, users can select any subgraph $G'_i \subseteq G_i$ in a view using their favorite selection mode. Similarly to a brushing interaction, they can then drag this selection across views to another view containing subgraph $G_j \subseteq {\cal{G}}$, and release the selection in it. Dragging the selection preserves the relative position of the selected elements which are only translated following the mouse position.

Upon release the selection $G'_i=(V'_i, E'_i)$, we check which nodes and edges belong to the target view's graph $G_j=(V_j, E_j)$, such as $v \in (V'_i \cap V_j)$ and $e \in (E'_j \cap E_j)$. Nodes and edges of the selection that do not satisfy this condition are faded out. Nodes and edges of the target view's graph that do not satisfy this condition are grayed. The position of nodes and edges that satisfy the condition are then gradually interpolated from their release position to their position in the target graph. This interpolation forms the animation.

Following Tversky's \emph{et al.} recommendations \cite{tversky2002animation} for control of animation, with the hold of a control key, we additionally provide a preview of animation to track individual elements (Fig. \ref{fig:drag-drop}(f)). In this mode this animation is executed on mouseover a target view, before the drop is released. The position of animation is then controlled by the mouse position thus the user can move backward or forward in the animation time to track individual elements.

\subsection{Implementation}

We implemented this interaction on top of d3 and HTML's drag and drop interaction events. But we used careful design to avoid conflict between all interactions. We want the drag and drop operate on a selection, but selections are usually implemented over a drag and drop interaction too (for example to create and brush a lasso selector). 

In order to achieve this, upon selection, we create a new SVG layer on top of the graph view's canvas, that is invisible to the user and recreate only the selected part of the graph. This SVG captures the drag interaction, and the user interaction actually drags this layer instead. Each other view's canvas also presents a temporary SVG upper layer which listens to the drop event. 

Finally, when the drop is called or the control key is pressed, the subgraph is copied at the drop position in the new layer and then we interpolate with animation to the positions as detailed in the previous section. The mouse position following an invisible line between the two views gives the percentage of progress of the animation. Therefore the user can advance to the full interpolation or go back in its early steps.

Note that color encoding at the end of interpolation is the same of the encoding from the target view. This is designed so users may use compare where the nodes were originated from in the previous view while helping them better track sub-elements of interest during the animation.

\section{Discussion and Conclusion}

We designed this drag and drop interaction in a highly multidimensional and multi-view context, for which linked was already serving different purposes, and more would have added confusion. The drag and drop interaction we introduce here is designed to support the analysis of dynamic communities. This is why we base our design on fixed choices such as the use of color encoding for categorical variables. Because our community structure is different for each of the small multiples, we cannot use the same encoding across all views. The animation becomes very helpful to match nodes and communities between views. 

Although drag and drop is an interaction limited by definition to two views only, it gives a nice ``physicality'' to the interaction, especially with the mouse control. Animation works well in that case because it is exactly between two views limited to a user-defined point-of-interest.  Informal feedbacks from users show promises in helping them engaging with the system. 

In this work, we limited ourselves the use of our drag and drop interaction to the \emph{comparison} task, but we only scratched the surface. The ``drop'' part actually captures the intent of a user (through the localization of the drop) and we have not exploited it yet. Small multiples are most appreciated from users \cite{van2013small} and when multiple separated views are provided, there is no doubt that drag and drop interactions make a great improvement so users may easily manipulate between data selections and data representations.

\bibliographystyle{abbrv-doi}

\bibliography{template}
\end{document}